\documentclass[aps,prl,twocolumn]{revtex4}

\usepackage{graphicx}

\usepackage{amssymb}
\usepackage{amsmath}
\usepackage{verbatim}
\usepackage{tabularx}

\usepackage{algorithmicx}
\usepackage{algpseudocode}

\begin{document}

\title{Self-Organized Criticality from Protected Mean-Field Dynamics:
Loop Stability and Internal Renormalization in Reflective Neural Systems}

\author{Byung Gyu Chae}

\affiliation{Electronics and Telecommunications Research Institute, 218 Gajeong-ro, Yuseong-gu, Daejeon 34129, Republic of Korea}

\begin{abstract}
The reflective homeostatic dynamics provides a minimal mechanism for
self-organized criticality in neural systems.
Starting from a reduced stochastic description, we demonstrate within the
MSRJD field-theoretic framework that fluctuation effects do not destabilize the
critical manifold.
Instead, loop corrections are dynamically regularized by homeostatic curvature,
yielding a protected mean-field critical surface that remains marginally stable
under coarse-graining.
Beyond robustness, we show that response-driven structural adaptation generates
intrinsic parameter flows that attract the system toward the vicinity of this surface without
external fine tuning.
Together, these results unify loop renormalization and adaptive response in a
single framework and establish a concrete route to autonomous criticality in
reentrant neural dynamics.
\end{abstract}

\maketitle

\emph{Introduction}---Modern neural architectures have achieved remarkable performance across a wide
range of cognitive and perceptual tasks. Nevertheless, their internal dynamics
remain largely feedforward or only weakly recurrent, limiting their capacity
for reflective computation, self-stabilization, and context-dependent
reinterpretation of internal states \cite{1,2,3,4,5,6,7}. In biological neural systems, by contrast,
reentrant signaling and homeostatic regulation are widely regarded as central
mechanisms supporting robust cognition, sensitivity to perturbations, and
long-term stability \cite{8,9,10,11,12}. Motivated by these considerations, we recently introduced
a class of \emph{reflective homeostatic networks}, in which neural activity is
recursively fed back into the computation and dynamically regulated by
homeostatic constraints \cite{13,14,15}.

When formulated in continuous time, these models admit a reduced dynamical
description in terms of a small number of collective variables. This
representation enables explicit analysis of stability, fluctuations, and
response properties \cite{16,17,18}. Initial investigations focused on local stability and
boundedness of activity, revealing a delicate balance between reentrant
amplification and nonlinear regulation. While this analysis establishes that
runaway dynamics can be avoided, it leaves open a more fundamental question \cite{10,19,20}:
\emph{how does the system organize its structural parameters globally, in the
presence of noise, without external fine tuning?}

Addressing this question requires moving beyond purely local considerations.
In particular, understanding the long-time and large-scale behavior of adaptive
reentrant systems demands a framework capable of treating fluctuations,
response, and structural evolution on equal footing \cite{21,22,23}. Remarkably, the stochastic
dynamics underlying reflective homeostatic networks admits a natural
field-theoretic formulation within the
Martin--Siggia--Rose--Janssen--De~Dominicis (MSRJD) formalism \cite{24,25,26}. This representation
provides a unified language in which conventional fluctuation-induced
renormalization and response-driven structural adaptation emerge from the same
effective action.

Within this framework, two distinct but complementary mechanisms become
apparent. On the one hand, Wilsonian coarse-graining over fast fluctuations
generates loop corrections that renormalize the effective couplings and govern
the stability of putative critical surfaces \cite{21,27}. On the other hand, the response
sector of the theory induces an intrinsic flow of structural parameters when
these parameters are allowed to evolve slowly in time \cite{28,29}. In this sense, the
response field plays a dual role: it encodes linear susceptibility to external
perturbations and simultaneously mediates an internal, adaptation-driven
renormalization of structure.

The central result of this work is that these two mechanisms conspire to drive
reflective homeostatic dynamics toward a \emph{self-organized critical state} \cite{30,31,32}.
Crucially, criticality here is not imposed by external tuning or boundary
conditions, but is dynamically selected by the internal coupling between
response, learning, and homeostatic regulation. While loop renormalization
determines the robustness of the critical surface against fluctuations,
response-weighted structural adaptation biases the dynamics toward the vicinity of that surface,
leading to an operating point at the edge of stability.

This mechanism differs fundamentally from traditional models of
self-organized criticality, which typically rely on slow driving and fast
dissipation or threshold-based avalanche dynamics \cite{30,32,33}. In the present setting,
criticality emerges as a functional consequence of reflective computation:
high susceptibility enhances adaptive pressure toward marginal stability, while
homeostatic nonlinearities suppress infrared instabilities and bound activity \cite{34,35,36,37}.
The resulting critical state is therefore both fluctuation-robust and
dynamically selected.

More broadly, our results suggest that criticality in adaptive systems may arise
not as a design principle, but as an emergent operating point enforced by
response-weighted renormalization. This perspective provides a bridge between
reentrant neural computation, field-theoretic methods, and self-organized
critical phenomena, and points toward a general mechanism by which complex
systems can maintain sensitivity and stability without external fine tuning.

\begin{figure}[t]
\centering
\includegraphics[scale=0.55, trim= 1.5cm 19cm 0cm 0cm]{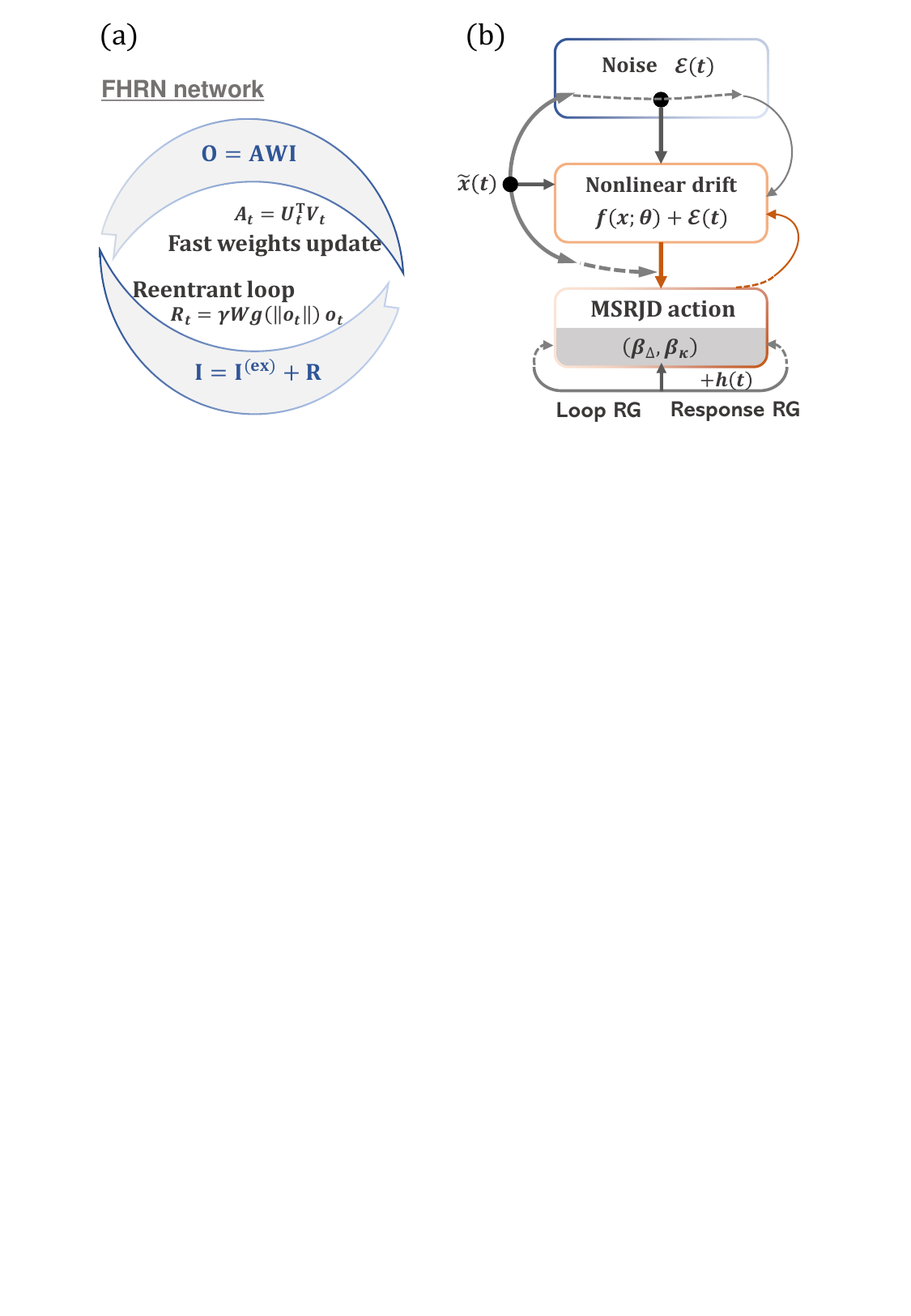}
\caption{Reflective homeostatic reentry and unified MSRJD description.
(a) Schematic of the reflective homeostatic reentrant network, in which reentrant amplification and fast-weight modulation confine the dynamics to a bounded activity shell.
(b) Corresponding MSRJD representation of the reduced stochastic dynamics, showing that both loop renormalization of fluctuations and response-driven structural adaptation are generated by the same action.
Together, these mechanisms yield a protected critical manifold that is dynamically selected without external fine tuning.
}
\label{fig:phase_portrait}
\end{figure}

\vspace{6pt}
\emph{Reflective homeostatic dynamics and reduced description}---Our starting point is the reflective homeostatic reentrant network (FHRN)
introduced in our earlier work in Fig 1(a), where reentrant activity is continuously
re-injected into the dynamics under the control of a global homeostatic nonlinearity \cite{14,15}.
At the microscopic level, the system is described by a continuous-time
dynamical equation for a $N$-dimensional state vector
$\mathbf{o}(t)\in\mathbb{R}^N$.
A defining structural property of the FHRN is rotational symmetry in state
space: the deterministic drift depends on $\mathbf{o}$ only through its
amplitude $r(t)=\|\mathbf{o}(t)\|$.

This symmetry allows the dynamics to be decomposed into radial and angular
sectors.
As shown previously, global homeostatic regulation renders the radial mode
strongly stable, while the angular degrees of freedom remain weakly constrained
and mix rapidly on the sphere.
As a result, the long-time and infrared behavior of the network is governed by
an effective radial dynamics, whereas angular fluctuations act as a source of
stochastic forcing.
Eliminating the angular modes under a controlled separation of time scales
therefore yields an effective one-dimensional stochastic equation for the
radial deviation from its steady-state value.

Concretely, after radial reduction and coarse-graining over angular dynamics,
the collective variable $r(t)$ obeys a stochastic differential equation of the
form
\begin{equation}
    \dot r(t) = f\!\big(r(t);\theta\big) + \xi(t),
    \label{eq:base_sde}
\end{equation}
where $\theta$ denotes a set of structural parameters inherited from the
microscopic model, and $\xi(t)$ is an effective Gaussian white noise arising
from the eliminated angular and microscopic degrees of freedom,
\begin{equation}
    \langle \xi(t) \rangle = 0,
    \qquad
    \langle \xi(t)\xi(t') \rangle = 2D\,\delta(t-t').
\end{equation}
Importantly, this noise term is not introduced phenomenologically but emerges
inevitably from the radial--angular decomposition and averaging procedure, as
detailed in Appendix~A.

The drift term $f(r;\theta)$ encodes both linear
instability induced by reentrant amplication and nonlinear saturation arising from
homeostatic regulation, $g(r)=\frac{1}{1+\kappa(r^2-1)}$.
The deterministic part of the radial flow admits a nontrivial fixed point
$r^\ast$, determined by the balance between reentrant amplification and
homeostatic curvature.
Expanding around this fixed point,
\begin{equation}
    x(t) \equiv r(t) - r^\ast ,
\end{equation}
the dynamics reduces to a local polynomial form,
\begin{equation}
    \dot x
    =
    a_1(\theta)\,x
    + a_2(\theta)\,x^2
    + a_3(\theta)\,x^3
    + \cdots
    + \xi(t),
    \label{eq:local_langevin}
\end{equation}
where the coefficients $a_n(\theta)$ are smooth functions of the underlying
structural parameters $\theta$, including the reentry gain and homeostatic
curvature (See the details in Appendix B).

A key result of the microscopic analysis is that the linear coefficient
\begin{equation}
    a_1(\lambda,\kappa)
    =
    -\frac{2(\kappa + \lambda - 1)}{\lambda}
\end{equation}
vanishes along a codimension-one manifold.
Introducing the deviation from criticality
\begin{equation}
    \Delta \equiv \lambda - 1 ,
\end{equation}
this marginal-stability manifold is given by $\kappa + \Delta = 0$.
Thus, the location of the dynamical critical surface emerges naturally
from the structure of the reflective homeostatic dynamics, rather than
being imposed by external tuning.

\vspace{6pt}
\emph{MSRJD formulation and renormalization structure}---To analyze fluctuation effects and adaptive dynamics on equal footing, we
formulate Eq. (4) within the
Martin--Siggia--Rose--Janssen--De~Dominicis (MSRJD) formalism \cite{24,25,26}. 
In this representation, the stochastic dynamics is expressed exactly as a path integral
over the state variable $x(t)$ and an auxiliary response field $\tilde x(t)$,
with action
\begin{equation}
\begin{aligned}
    S[x,\tilde x;\theta]
    &=
    \int dt\;
    \tilde x(t)\big(\dot x(t)-f(x(t);\theta)\big)
    -
    D\int dt\;\tilde x(t)^2 .
\end{aligned}
\label{eq:msrjd_action}
\end{equation}
The MSRJD action encodes both the deterministic drift and stochastic
fluctuations, and serves as the starting point for all subsequent analytical
developments.

A central advantage of the MSRJD formulation is that it exposes the structural
role of fluctuations and response within a single theoretical framework, as illustrated in Fig. 1(b).
Notably, the same action supports two distinct but complementary
renormalization procedures. First, conventional Wilsonian coarse-graining over
fast fluctuations of $x$ and $\tilde x$ generates loop corrections that
renormalize the effective couplings $a_1$ and $a_3$, thereby determining the
stability of the dynamical critical surface under noise. Second, when the structural
parameters $\theta$ are allowed to evolve slowly in time, the response sector of
the action induces an intrinsic flow of couplings driven by sensitivity and
susceptibility.

In this sense, the response field $\tilde x$ plays a dual role. It governs the
linear response of the system to external perturbations, and simultaneously
mediates an internal renormalization of structure when parameters are adaptive.
This unified perspective allows us to disentangle two logically distinct
questions. Loop renormalization determines whether a critical manifold, once approached, is stable against fluctuations,
while response-driven parameter flow
determines whether the dynamics is attracted toward the vicinity (or edge) of that manifold in the absence of external fine tuning.
As we show in the following sections, reflective
homeostatic dynamics realizes both mechanisms within the same theoretical
framework, leading to a critical state that is both fluctuation-robust and
dynamically selected.

\vspace{6pt}
\emph{Loop renormalization and protected mean-field criticality}—We examine the stability of the critical manifold under stochastic fluctuations
by performing a controlled Wilsonian loop expansion of the MSRJD action, as detailed in Appendix~C.
The central issue is whether the critical surface identified at the
deterministic mean-field level survives once fluctuations are integrated out,
or whether it is destabilized by infrared divergences, as often occurs in
low-dimensional interacting systems.

To this end, we linearize the drift term around the origin,
\begin{equation}
f(x;\theta) \approx a_1 x ,
\end{equation}
and decompose the action into a Gaussian part and an interaction,
\begin{equation}
S = S_0 + S_{\mathrm{int}} ,
\end{equation}
with
\begin{equation}
\begin{aligned}
    S_0
    &=
    \int dt\;
    \tilde x(t)\big(\dot x(t)-a_1 x(t)\big)
    -
    D\int dt\;\tilde x(t)^2 ,
    \\
    S_{\mathrm{int}}
    &=
    \int dt\;
    \tilde x(t)\big(-a_2 x(t)^2-a_3 x(t)^3\big).
\end{aligned}
\label{eq:action_split}
\end{equation}

The Gaussian theory defined by $S_0$ yields the bare response and correlation
functions,
\begin{equation}
G_R(\omega)
=
\frac{1}{-i\omega + a_1},
\qquad
G_C(\omega)
=
\frac{2D}{\omega^2 + a_1^2},
\end{equation}
where $G_R(\omega)=\langle x(\omega)\tilde x(-\omega)\rangle_0$ and
$G_C(\omega)=\langle x(\omega)x(-\omega)\rangle_0$.
As the critical surface $a_1 \to 0$ is approached, the static susceptibility
diverges as $\chi \sim |a_1|^{-1}$, reflecting the growing temporal correlations
characteristic of critical slowing down.

At the level of naive perturbation theory, the leading fluctuation correction
generated by the cubic interaction takes the form
$
\delta a_1
\sim
- a_3
\int \frac{d\omega}{2\pi}\, G_C(\omega),
$
which diverges as $\delta a_1 \sim - a_3 D/|a_1|$ in the limit $a_1 \to 0$.
Taken at face value, this infrared divergence would suggest that fluctuations
destabilize the critical surface, as depicted in  Fig. 2(a).

However, this conclusion is premature, as it ignores the Wilsonian structure of
the renormalization procedure.
In a Wilsonian RG step, only fast modes in a thin frequency shell
$\Lambda/b < |\omega| < \Lambda$ are integrated out.
At one-loop order, the resulting correction to the linear coefficient is
\begin{equation}
\delta a_1
=
- 3 a_3
\int_{\Lambda/b<|\omega|<\Lambda}
\frac{d\omega}{2\pi}\, G_C(\omega)
+ \cdots ,
\end{equation}
where the numerical prefactor arises from causal contractions.
For a thin shell, this integral evaluates to
\begin{equation}
\int_{\Lambda/b<|\omega|<\Lambda}
\frac{d\omega}{2\pi}\, G_C(\omega)
\simeq
\frac{2D}{\pi}
\frac{\Lambda}{\Lambda^2 + a_1^2}\, d\ell ,
\end{equation}
which remains finite at any fixed RG scale.

The apparent infrared divergence instead emerges only when the RG flow is
iterated toward the critical surface.
Crucially, in reflective homeostatic dynamics the cubic coupling $a_3$ is not an
independent parameter.
Rather, it encodes a curvature constraint that becomes increasingly effective
as the activity amplitude grows.
As illustrated in Fig.~2(b), this nonlinear saturation generates a finite
characteristic amplitude scale,
$
x_{\mathrm{typ}}^2 \sim -a_1/a_3,
$
which acts as a self-consistent infrared cutoff.
As a result, the same homeostatic mechanism that bounds trajectories at the
deterministic level also regularizes the infrared sector of the field theory,
preventing runaway fluctuation-induced instabilities, in Fig. 2(c).

Collecting the tree-level scaling with the one-loop correction and expressing
the result in terms of its infrared scaling structure, the flow of the linear
coupling can be summarized by an effective beta function of the form
\begin{equation}
\beta_{a_1}
:=
\frac{d a_1}{d\ell}
\sim
- a_1
+ c\,\frac{a_3 D}{|a_1|}
+ \cdots ,
\label{eq:beta_a1}
\end{equation}
where $\ell$ denotes the logarithmic coarse-graining scale and $c>0$ is a
nonuniversal constant.
This expression should be understood as an infrared scaling form rather than an
exact Wilsonian beta function; while subleading terms depend on the
regularization scheme, the qualitative structure of the flow is robust.

As a consequence, the dynamical condition $a_1=0$, corresponding to a locally
marginal relaxation rate, is neither shifted nor destroyed by loop corrections,
but instead remains marginally stable.
Mean-field scaling is preserved: fluctuations renormalize amplitudes but do not
generate anomalous critical exponents.
We therefore refer to this regime as a \emph{protected mean-field critical
manifold}.

While the loop expansion is naturally expressed in terms of the local normal-form
coefficients $(a_1,a_3)$, the underlying structural control parameters of the
reflective dynamics are the reentry gain $\lambda$ and the homeostatic curvature
$\kappa$.
Near the critical manifold, these parameters are related by Eq. (5),
so that the vanishing of the local linear coefficient $a_1$
corresponds to the condition $\kappa+\Delta=0$ in the underlying
structural parameter space.
The loop-induced renormalization of $a_1$ can thus be reinterpreted as a
renormalization of the deviation from criticality $\Delta$.

After absorbing scheme-dependent constant shifts into a rescaling of time, the
Wilsonian flow assumes a universal normal form.
Introducing a dimensionless deviation $\tilde\Delta$, defined by evaluating
nonuniversal prefactors on a fixed reference slice of the structural parameter
space (e.g.\ $\kappa=1$), the resulting beta
function reads
\begin{equation}
    \beta_{\tilde\Delta}
    =
    \tilde\Delta
    -
    \tilde\Delta^{\,2}
    +
    \mathcal O(\tilde\Delta^{\,3}),
    \label{eq:beta_delta_nf}
\end{equation}
demonstrating that the critical surface remains infrared attractive and
marginally stable, independent of the details of infrared regularization.

The flow of the homeostatic curvature $\kappa$ is induced by the renormalization
of the cubic response vertex, which controls the effective curvature of the
radial restoring force.
In our scheme, $\kappa$ is defined through the renormalized cubic coefficient
$a_3^R$, so that fluctuations renormalizing the $-\tilde x x^3$ interaction
directly translate into a flow of $\kappa$.
At one loop, mixed and self-interaction diagrams involving the quadratic and
cubic vertices generate corrections to $a_3$, and, through this identification,
induce a curvature flow driven by critical fluctuations (See the details in Appendix D).

As a result, the beta function for $\kappa$ takes the generic normal-form
structure
\begin{equation}
    \beta_{\kappa}
    =
    \alpha_1(\kappa-\kappa_0)
    +
    \alpha_2\,\tilde\Delta
    +
    \alpha_3\,\tilde\Delta^{\,2}
    +
    \cdots ,
    \label{eq:beta_kappa_nf}
\end{equation}
where the coefficients $\alpha_i$ are determined by explicit one-loop vertex
diagrams and depend on the microscopic noise strength and cutoff.
On the critical manifold $\tilde\Delta=0$, the curvature flows toward a finite
fixed value $\kappa_0$, ensuring bounded radial fluctuations and stabilizing the
reflective shell.

Taken together, these results establish that loop renormalization preserves the
existence and structure of the reflective critical manifold, but does not by
itself select a unique operating point along it.
In the following section, we show that this remaining degeneracy is lifted
dynamically: response-driven structural adaptation generates an intrinsic flow
in $(\Delta,\kappa)$ space that actively attracts the system toward a specific
point on this protected manifold, giving rise to self-organized criticality
without external fine tuning.

\begin{figure}[t]
\centering
\includegraphics[scale=0.48, trim= 0.7cm 20cm 0cm 0cm]{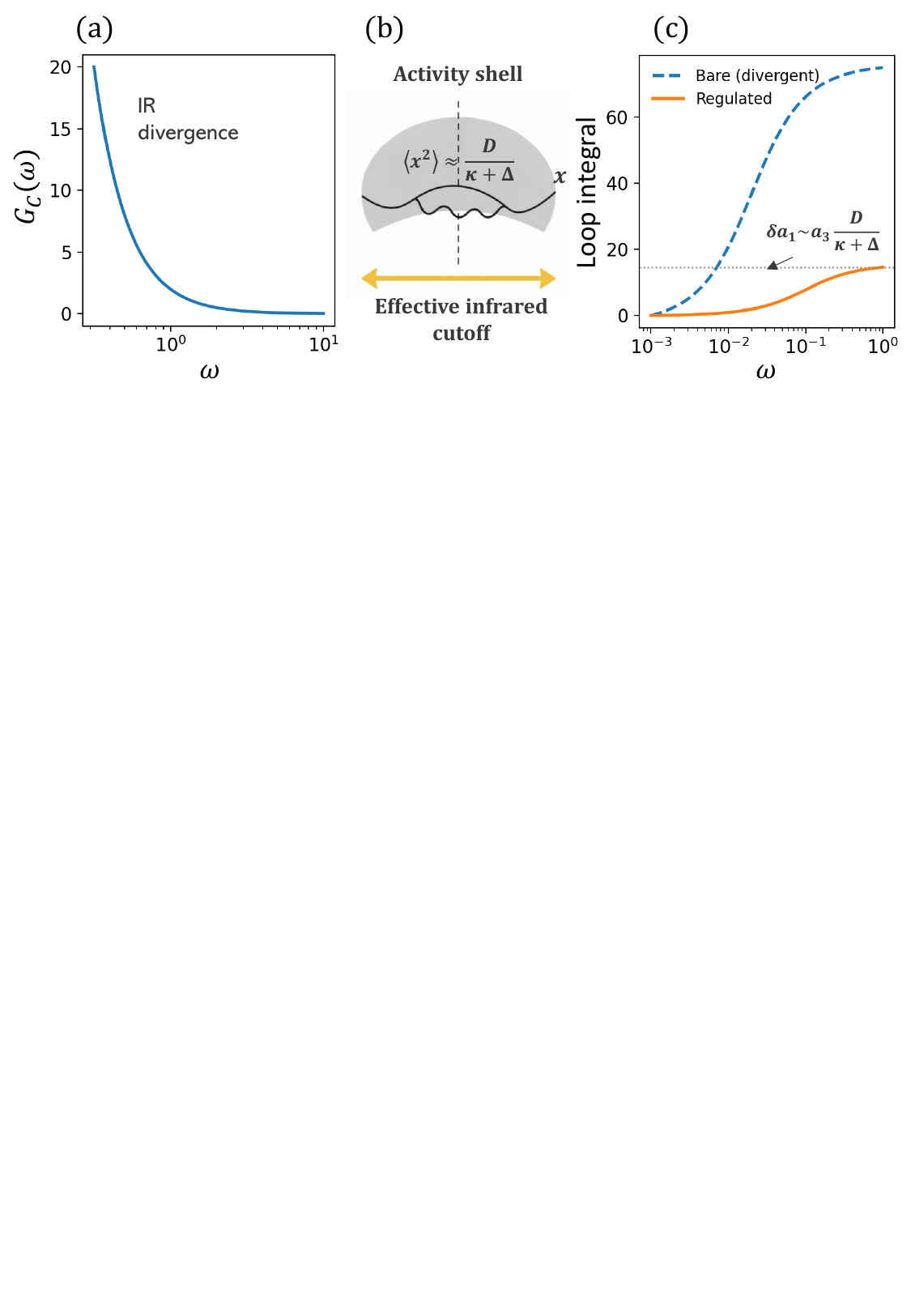}
\caption{Infrared regularization by homeostatic curvature.
(a) Bare correlation propagator $G_C(\omega)$ of the Gaussian theory,
which diverges in the infrared as $a_1\to 0$.
(b) Nonlinear saturation encoded in the curvature $a_3$ generates a finite
characteristic amplitude scale,
$
x_{\mathrm{typ}}^{2}\sim -a_1/a_3,
$
which provides an effective infrared cutoff.
(c) As a result, the loop integral contributing to $\delta a_1$ is regulated
self-consistently, preventing destabilization of the critical surface.
}
\label{fig:S2}
\end{figure}

\vspace{6pt}
\emph{Response-driven structural dynamics}---The loop analysis establishes that the critical surface
is robust once it is reached. We now address a distinct and complementary
question: \emph{how does the system arrive at this surface in the absence of external
fine tuning?} To answer this, we explicitly incorporate external perturbations
and slow structural adaptation into the reduced description.

The reduced dynamics takes the form
\begin{equation}
    \dot x(t) = f\!\big(x(t);\theta\big) + h(t) + \xi(t),
    \label{eq:reduced_langevin}
\end{equation}
where $h(t)$ is a slowly varying external input or effective low-frequency
drive, representing either environmental perturbations or internally
generated reentrant signals.
Retaining the leading nonlinearities compatible with
bounded activity, the drift admits a generic normal form,
$ f(x;\theta) = a_1(\theta)\,x - a_3(\theta)\,x^3 + \mathcal O(x^5)$.

To characterize how the system responds to external inputs, we again employ the
MSRJD formalism, now retaining the explicit dependence on the source field $h(t)$. 
A defining feature of this representation is that the response field
$\tilde x(t)$ is conjugate to the external input.
Functional differentiation with respect to $h(t)$ yields the linear response
kernel,
\begin{equation}
    R(t,t')
    :=
    \frac{\delta\langle x(t)\rangle}{\delta h(t')}
    \bigg|_{h=0}
    =
    \langle x(t)\,\tilde x(t')\rangle .
    \label{eq:linear_response}
\end{equation}
The response field thus provides a direct measure of the system's sensitivity to
external perturbations, which becomes strongly enhanced near the dynamically marginal regime
($a_1\to 0$) identified in the loop analysis.

So far, the structural parameters $\theta$ have been treated as fixed.
In adaptive reentrant systems, however, these parameters evolve slowly in
response to activity and input.
To capture this effect, we allow $\theta$ to vary on a time scale much longer
than that of the fast dynamics of $x(t)$,
\begin{equation}
    \tau_\theta \gg \tau_x .
\end{equation}
Within this separation of scales, the fast variables $(x,\tilde x)$ equilibrate
quasi-statically for fixed $\theta$, while the structural parameters evolve under
averaged response-driven forces.

A natural and minimal assumption, consistent with the MSRJD formulation,
is that the slow evolution of $\theta$ follows the gradient of the action,
\begin{equation}
    \dot\theta
    =
    -\eta\,\frac{\delta S}{\delta \theta},
    \qquad \eta>0,
    \label{eq:theta_flow_def}
\end{equation}
which yields the response-weighted adaptation rule
\begin{equation}
    \dot\theta
    =
    \eta\;
    \tilde x(t)\,
    \partial_\theta f\!\big(x(t);\theta\big).
    \label{eq:learning_rule}
\end{equation}
Averaging over the fast stochastic dynamics produces an effective deterministic
flow in parameter space,
\begin{equation}
    \langle \dot\theta\rangle
    =
    \eta\;
    \big\langle
    \tilde x\,\partial_\theta f(x;\theta)
    \big\rangle .
    \label{eq:theta_flow_avg}
\end{equation}

This flow is qualitatively distinct from conventional loop renormalization.
While loop corrections describe how fluctuations renormalize couplings at fixed
structure, Eq.~\eqref{eq:theta_flow_avg} describes how the structure itself is
dynamically reshaped by response and sensitivity.

\vspace{6pt}
\emph{Response-driven flows toward the critical manifold}---We now specialize the general response-driven framework 
to the radial dynamics associated with reflective homeostatic reentry.
As established in prior work \cite{14}, the reduced radial flow takes the form
\begin{equation}
    \dot r
    =
    f(r;\lambda,\kappa)
    =
    r\Big[-1+\frac{\lambda}{1+\kappa(r^2-1)}\Big],
    \label{eq:radial_drift}
\end{equation}
where $\lambda$ controls the effective reentry gain and $\kappa>0$ sets the
strength of global homeostatic curvature.
We define the distance to criticality as $\Delta = \lambda-1$.
In the absence of noise and external input, the nontrivial fixed point of
Eq.~\eqref{eq:radial_drift} satisfies
\begin{equation}
    r_\ast^2
    =
    1+\frac{\Delta}{\kappa},
    \label{eq:shell_radius}
\end{equation}
so that the critical manifold $\Delta=0$ corresponds to a shell of unit radius,
$r_\ast=1$.

To quantify the response entering the structural flows, we linearize the radial
dynamics around the shell, $r(t)=r_\ast+\delta r(t)$.
To leading order one obtains
\begin{equation}
    \delta\dot r
    =
    a(\lambda,\kappa)\,\delta r + h(t) + \xi(t),
    \label{eq:ou_form}
\end{equation}
with
\begin{equation}
    a(\lambda,\kappa)
    =
    \partial_r f(r;\lambda,\kappa)\big|_{r=r_\ast}
    =
    -\frac{2\kappa r_\ast^2}{\lambda}.
    \label{eq:stability_coeff}
\end{equation}
For $\kappa>0$, the shell is linearly stable.
The resulting Ornstein--Uhlenbeck process yields a stationary variance
\begin{equation}
    \mathrm{Var}[r]
    =
    \langle (\delta r)^2\rangle
    =
    \frac{D_{\mathrm{tot}}}{-a(\lambda,\kappa)},
    \label{eq:radial_variance}
\end{equation}
where $D_{\mathrm{tot}}$ includes both intrinsic noise and the effective power
of slow external drive (See the details in Appendix E).
Near criticality ($\lambda\approx 1$), this reduces to
\begin{equation}
    \mathrm{Var}[r]
    \approx
    \frac{D_{\mathrm{tot}}}{2(\kappa+\Delta)}.
    \label{eq:variance_critical}
\end{equation}
Equation~\eqref{eq:variance_critical} highlights the dual role of $\kappa$:
it stabilizes the shell while simultaneously controlling its fluctuation
thickness.

We first consider the slow adaptation of the reentry gain $\lambda$.
As discussed in the preceding section, the response-weighted learning rule yields
\begin{equation}
    \langle \dot\lambda\rangle
    =
    \eta\,
    \big\langle
    \tilde r\,\partial_\lambda f(r;\lambda,\kappa)
    \big\rangle .
    \label{eq:lambda_flow_general}
\end{equation}
A minimal and physically transparent closure is to assume that adaptation acts
to reduce deviations of the shell from the critical radius $r=1$ (See the details in Appendix F).
Under time-scale separation, the fast dynamics concentrates near $r_\ast$,
so that
\begin{equation}
    \big\langle r^2-1 \big\rangle
    \approx
    r_\ast^2-1
    =
    \frac{\Delta}{\kappa}.
\end{equation}
Consequently, the induced slow flow for $\Delta=\lambda-1$ takes the form
\begin{equation}
    \langle \dot\Delta\rangle
    \approx
    -c_\Delta\,\eta\,\frac{\Delta}{\kappa},
    \qquad c_\Delta>0,
    \label{eq:Delta_flow_final}
\end{equation}
implying that $\Delta=0$ is an attractive fixed point of the response-driven
adaptation dynamics for any $\kappa>0$.
Thus the reentry gain is naturally tuned toward the critical manifold by
sensitivity-weighted structural adaptation.

We next consider the adaptation of the homeostatic curvature $\kappa$.
Since $\kappa$ controls the sensitivity of the gain to radial excursions, its
slow evolution regulates the thickness of the radial shell.
A minimal closure consistent with the MSRJD formulation is to drive $\kappa$
toward a target fluctuation level $\sigma_{\mathrm{tar}}^2$,
\begin{equation}
    \langle \dot\kappa\rangle
    \propto
    \eta\,
    \big(
    \mathrm{Var}[r]
    -
    \sigma_{\mathrm{tar}}^2
    \big).
    \label{eq:kappa_flow_general}
\end{equation}
Substituting the variance \eqref{eq:variance_critical} yields
\begin{equation}
    \langle \dot\kappa\rangle
    \approx
    c_\kappa\,\eta
    \Big(
    \frac{D_{\mathrm{tot}}}{2(\kappa+\Delta)}
    -
    \sigma_{\mathrm{tar}}^2
    \Big),
    \qquad c_\kappa>0.
    \label{eq:kappa_flow_final}
\end{equation}
When fluctuations exceed the target level, $\kappa$ increases and tightens the
shell, whereas insufficient fluctuations lead to a decrease of $\kappa$.
This adaptive feedback suppresses runaway amplification while maintaining
sensitivity.

The coupled slow flows \eqref{eq:Delta_flow_final} and
\eqref{eq:kappa_flow_final} admit a stable fixed point given by
\begin{equation}
    \Delta^\ast = 0,
    \qquad
    \kappa^\ast
    =
    \frac{D_{\mathrm{tot}}}{2\sigma_{\mathrm{tar}}^2}.
    \label{eq:critical_fixed_point}
\end{equation}
At this point, the system operates at the edge of stability ($r_\ast=1$) while
maintaining a finite fluctuation thickness set by the balance between stochastic
forcing and homeostatic regulation.
Importantly, this operating point is reached without fine tuning of parameters:
it emerges dynamically from response-driven adaptation under generic conditions.

Figure 3 illustrates the slow structural flows in the $(\Delta,\kappa)$ plane induced by response-driven adaptation.
When only marginal loop renormalization is retained, trajectories are attracted toward the protected critical manifold 
$\Delta=0$, forming a line attractor with no unique operating point.
Including response-weighted structural dynamics lifts this degeneracy: the combined action of susceptibility-driven attraction 
and homeostatic stabilization focuses the flow toward a finite point on the critical manifold.
Together, these flows demonstrate how reflective homeostatic dynamics self-organize toward criticality without external fine tuning.

\begin{figure}[t]
\centering
\includegraphics[scale=0.47, trim= 0.2cm 19.5cm 0cm 0cm]{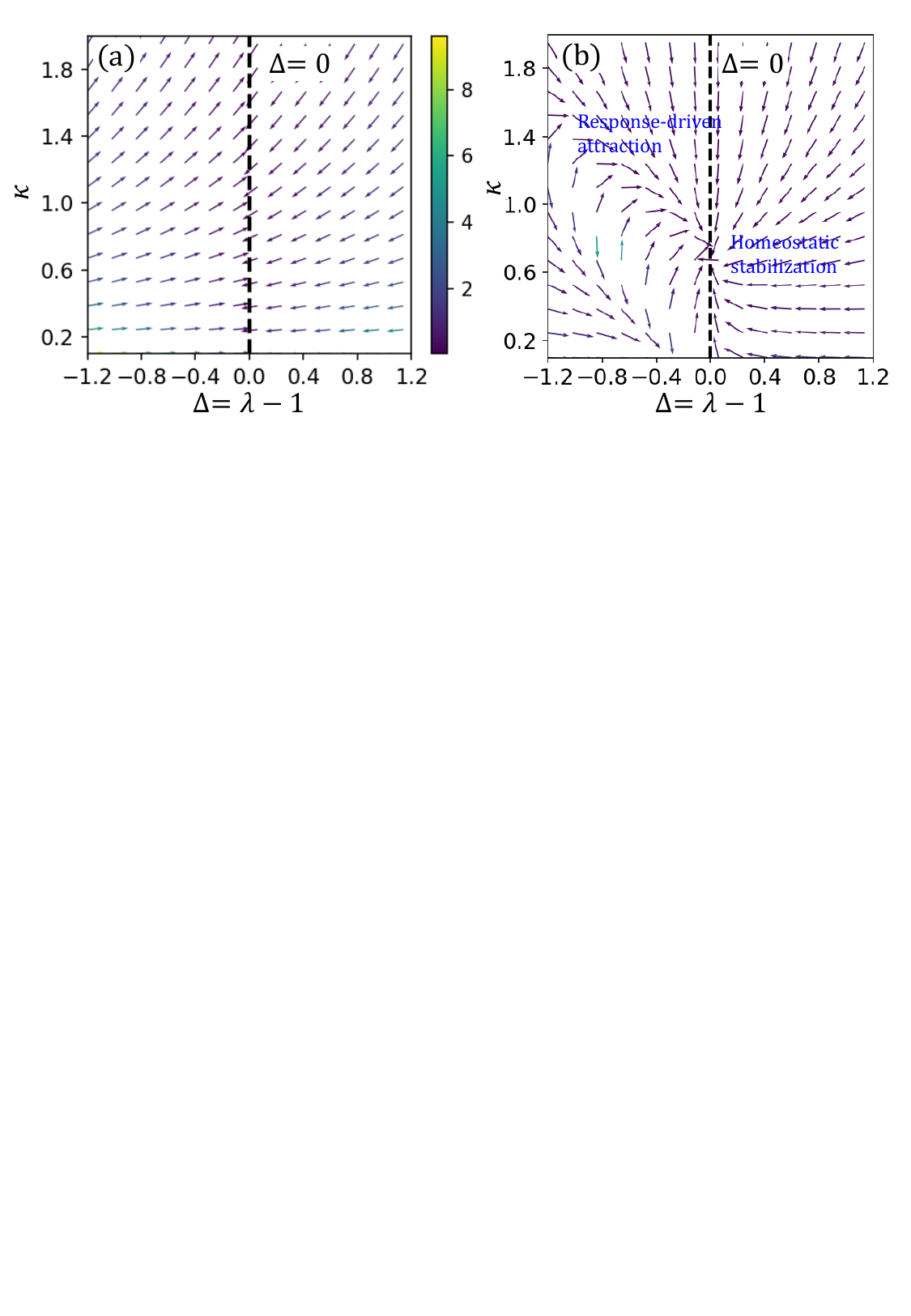}
\caption{Response-driven flows in $(\Delta,\kappa)$ space.
(a) Loop renormalization preserves a marginally stable critical manifold at $\Delta=0$.
(b) Response-driven structural adaptation lifts this degeneracy and dynamically selects a unique edge-of-criticality fixed point, realizing self-organized criticality.
}
\label{fig:S2}
\end{figure}

\vspace{6pt}
\emph{Discussions}---The central result of this work is the identification of two logically distinct,
yet dynamically intertwined, mechanisms governing criticality in reflective
homeostatic systems.
On the one hand, conventional loop renormalization within the MSRJD field theory
demonstrates that the mean-field critical surface is \emph{protected} against
infrared destabilization.
On the other hand, response-driven structural dynamics actively \emph{selects}
this critical surface as an attractive operating point, without the need for
external tuning.

These two roles are conceptually orthogonal.
Loop renormalization addresses a question of \emph{stability}: given a critical
manifold at the mean-field level, do fluctuations destroy it or shift it?
By contrast, response-driven flows address a question of \emph{emergence}: why
should the system evolve toward this manifold in the first place?
The present framework shows that reflective homeostatic dynamics realizes both
robustness and selection within a single theoretical structure.

It is instructive to compare the present mechanism with classical notions of
self-organized criticality (SOC) \cite{30,31,32}.
In sandpile models and related systems, criticality emerges through slow
driving and dissipation, typically relying on conservation laws or explicit
separation between driving and relaxation scales.
In such models, critical behavior is often identified retrospectively through
scaling laws, while the underlying dynamical selection mechanism remains implicit.

By contrast, in the reflective homeostatic framework studied here, criticality
emerges through \emph{internal response-driven adaptation}.
Structural parameters evolve according to the system's own susceptibility and
response, encoded explicitly in the MSRJD response field.
As a result, attraction to criticality is not imposed externally but arises as
a dynamical consequence of reentrant amplification combined with homeostatic
regulation.
This places the present mechanism closer to a form of \emph{adaptive SOC},
in which control parameters themselves evolve dynamically via internal response rather than external driving..

The preservation of mean-field scaling despite the presence of fluctuations is
a nontrivial feature of the present theory.
Ordinarily, one would expect infrared divergences to renormalize critical
exponents in low-dimensional systems.
Here, however, the same nonlinearities that stabilize activity amplitudes also
act as a self-consistent infrared regulator.
This leads to a scenario in which fluctuations renormalize amplitudes and
response strengths but leave the qualitative structure of the critical manifold
intact.

Importantly, this protection does not rely on symmetry, dimensionality, or
fine-tuning.
Instead, it emerges dynamically from the interplay between reentrant gain and
homeostatic curvature.
This mechanism may therefore be relevant to a broader class of adaptive
recurrent systems beyond the specific model studied here.

From a computational perspective, the results suggest a principled route by
which reentrant systems can operate near maximal sensitivity while remaining
stable.
The critical manifold corresponds to a regime of enhanced responsiveness,
long correlation times, and flexible amplification, while homeostatic regulation
ensures bounded activity.
Such properties are widely associated with efficient information processing and
adaptive behavior in biological neural systems.

More broadly, the present framework provides a concrete field-theoretic bridge
between dynamical systems models of recurrence and statistical descriptions of
criticality.
By embedding learning and adaptation directly into the response sector of the
theory, it offers a route toward unifying stability analysis, fluctuation theory,
and self-organization within a single mathematical language.

\vspace{6pt}
\emph{Conclusion}---We have shown that reflective homeostatic dynamics provides a minimal and self-consistent mechanism for the emergence of self-organized criticality in reentrant systems. 
Starting from a reduced stochastic description, 
we demonstrated within the MSRJD field-theoretic framework that fluctuation effects do not destabilize the critical manifold; instead, 
loop corrections are dynamically regularized by homeostatic curvature, yielding a protected mean-field critical surface that remains marginally stable under coarse-graining. 
Beyond robustness, we identified a complementary mechanism whereby response-driven structural adaptation generates intrinsic flows of the control parameters, 
attracting the system toward this critical manifold without external fine-tuning. 
The resulting operating point lies at the edge of stability, characterized by bounded activity and finite fluctuations set by the balance between stochastic forcing and homeostatic regulation. 
Together, these results establish reflective homeostatic systems as a concrete class 
in which robustness to fluctuations and dynamic self-organization toward criticality arise from the same underlying structure, 
unifying loop renormalization and adaptive response within a single theoretical framework.

\vspace{6pt}
\emph{Acknowledgements}---This work was partially supported by the Institute of Information \& Communications Technology Planning \& Evaluation (IITP) grant 
funded by the Korea government (MSIT) (IITP-RS-2025-02214780).

The author acknowledges the support of ChatGPT (GPT-5, OpenAI) for assistance in literature review and conceptual structuring during early development.

\clearpage
\appendix

\renewcommand{\thefigure}{S\arabic{figure}}
\renewcommand{\theequation}{S\arabic{equation}}

\setcounter{figure}{0}
\setcounter{equation}{0}

\vspace*{1.5cm}
{\centering\large\bfseries End Matter\par}
\vspace{1.0cm}

\section{Appendix A: Elimination of angular degrees of freedom and emergence of stochastic radial dynamics}

We consider a deterministic dynamical system in $\mathbb{R}^N$,
\begin{equation}
    \dot{\mathbf{x}}(t) = \mathbf{F}(\mathbf{x}),
    \qquad \mathbf{x}\in\mathbb{R}^N ,
\end{equation}
which represents the full reentrant dynamics prior to coarse-graining.
In the FHRN construction, $\mathbf{F}$ includes the reentrant operator,
nonlinear activation, and global homeostatic normalization.

\vspace{10pt}
\emph{(I) Radial--angular decomposition\rm{:}}
We introduce polar coordinates in state space,
\begin{equation}
    \mathbf{x}(t) = r(t)\,\mathbf{n}(t),
    \qquad
    r(t) = \|\mathbf{x}(t)\|,
    \qquad
    \|\mathbf{n}(t)\| = 1 .
\end{equation}
Taking a time derivative yields the identity
\begin{equation}
    \dot{\mathbf{x}}
    =
    \dot r\,\mathbf{n}
    +
    r\,\dot{\mathbf{n}} .
\end{equation}

Projecting the dynamics parallel and orthogonal to $\mathbf{n}$ gives
\begin{align}
    \dot r
    &=
    \mathbf{n}\cdot \mathbf{F}(r\mathbf{n}),
    \label{eq:radial_exact}
    \\[6pt]
    r\,\dot{\mathbf{n}}
    &=
    \mathbf{F}(r\mathbf{n})
    -
    \big(\mathbf{n}\cdot \mathbf{F}(r\mathbf{n})\big)\mathbf{n}.
    \label{eq:angular_exact}
\end{align}

\vspace{10pt}
\emph{(II) Isotropy and angular averaging\rm{:}}
A defining structural assumption of the FHRN dynamics is isotropy:
\begin{equation}
    \mathbf{F}(R\mathbf{x}) = R\,\mathbf{F}(\mathbf{x}),
    \qquad \forall R \in O(N).
\end{equation}
Under this condition, the radial projection admits a decomposition
\begin{equation}
    \mathbf{n}\cdot \mathbf{F}(r\mathbf{n})
    =
    F_r(r) + \delta F(r,\mathbf{n}),
\end{equation}
where
\begin{equation}
    F_r(r)
    :=
    \langle \mathbf{n}\cdot \mathbf{F}(r\mathbf{n}) \rangle_{\mathbf{n}}
\end{equation}
denotes the angular average over the unit sphere, and
\begin{equation}
    \langle \delta F(r,\mathbf{n}) \rangle_{\mathbf{n}} = 0 .
\end{equation}

Thus, the exact radial equation can be written as
\begin{equation}
    \dot r = F_r(r) + \delta F(r,\mathbf{n}(t)).
    \label{eq:radial_with_fluctuations}
\end{equation}

\vspace{10pt}
\emph{(III) Time-scale separation and mixing of angular dynamics\rm{:}}
Global homeostatic normalization ensures that the radial mode is
strongly stabilized,
\begin{equation}
    r(t) \xrightarrow{t\to\infty} r_\ast ,
\end{equation}
with a characteristic relaxation time $\tau_r$.
In contrast, the angular dynamics \eqref{eq:angular_exact} remain
weakly constrained and generically exhibit fast mixing on the sphere.
We assume a clear separation of time scales,
\begin{equation}
    \tau_n \ll \tau_r ,
\end{equation}
where $\tau_n$ is the angular correlation time.

For $|t-t'|\gg\tau_n$, angular correlations decay,
\begin{equation}
    \big\langle n_i(t)\,n_j(t') \big\rangle
    \;\longrightarrow\;
    \frac{1}{N}\,\delta_{ij}.
\end{equation}

\vspace{10pt}
\emph{(IV) Averaging theorem and emergence of stochastic forcing\rm{:}}
Equation \eqref{eq:radial_with_fluctuations} describes a slow variable
$r(t)$ driven by a rapidly fluctuating term $\delta F(r,\mathbf{n}(t))$
with zero mean and finite variance.
Standard averaging results for deterministic systems with fast chaotic
or mixing degrees of freedom imply that, in the limit
$\tau_n/\tau_r \to 0$,
\begin{equation}
    \delta F(r,\mathbf{n}(t))
    \;\Longrightarrow\;
    \xi(t),
\end{equation}
where $\xi(t)$ is an effective Gaussian white noise.
Collecting the above results, the coarse-grained radial dynamics take
the form
\begin{equation}
    \dot r = F_r(r) + \xi(t).
\end{equation}

\section{Appendix B: Fixed-point expansion and local polynomial coefficients}

We consider the radial (scalar) stochastic ODE
\begin{equation}
  \dot r(t) = f(r(t)) + \xi(t),
  \qquad
  f(r) := F(r;\lambda,\kappa)\,r,
  \label{eq:r_ode}
\end{equation}
with
\begin{equation}
  F(r;\lambda,\kappa)
  :=
  -1+\frac{\lambda}{1+\kappa(r^2-1)},
  \quad
  \Delta := \lambda-1,
  \: 
  \kappa>0.
  \label{eq:F_def}
\end{equation}
Define the auxiliary denominator
\begin{equation}
  D(r) := 1+\kappa(r^2-1),
  \quad\Rightarrow\quad
  F(r)=-1+\lambda\,D(r)^{-1}.
  \label{eq:D_def}
\end{equation}

Assume a Gaussian white noise with
\begin{equation}
  \langle \xi(t)\rangle = 0,
  \qquad
  \langle \xi(t)\xi(t')\rangle = 2D\,\delta(t-t'),
  \label{eq:noise}
\end{equation}
where $D$ is the noise strength.

We use
\begin{equation}
  x(\omega)=\int_{-\infty}^{\infty}dt\,e^{i\omega t}x(t),
  \qquad
  x(t)=\int\frac{d\omega}{2\pi}\,e^{-i\omega t}x(\omega),
  \label{eq:fourier_conv}
\end{equation}
so that $\partial_t \leftrightarrow -i\omega$.

\vspace{10pt}
\emph{(I) Fixed point $r_*$ and local variable $x=r-r_*$\rm{:}}
A nontrivial (``active shell'') fixed point with $r_*\neq 0$ is defined by
\begin{equation}
  \dot r=0
  \quad\Leftrightarrow\quad
  f(r_*) = F(r_*)r_* = 0
  \quad\Leftrightarrow\quad
  F(r_*)=0.
  \label{eq:fixedpoint_condition}
\end{equation}
Using \eqref{eq:F_def}--\eqref{eq:D_def},
\begin{align}
  F(r_*)=0
  \ &\Longleftrightarrow\ 
  -1+\frac{\lambda}{D(r_*)}=0
  \nonumber\\
  &\Longleftrightarrow\
  r_*^2=1+\frac{\Delta}{\kappa}.
  \label{eq:rstar}
\end{align}
Hence if $\Delta>0$ and $\kappa>0$ then $r_*^2>1$ and $r_*>1$.

Define the local fluctuation variable
\begin{equation}
  x(t):=r(t)-r_*
  \qquad\Rightarrow\qquad
  r(t)=r_*+x(t).
  \label{eq:x_def}
\end{equation}
Then \eqref{eq:r_ode} becomes
\begin{equation}
  \dot x(t) = f(r_*+x(t)) + \xi(t),
  \label{eq:x_ode_exact}
\end{equation}
since $\dot r=\dot x$ and $r_*$ is constant.

\vspace{10pt}
\emph{(II) Taylor expansion of the drift around $r_*$\rm{:}}
Expand $f(r)$ in Taylor series at $r=r_*$:
\begin{equation}
  f(r_*+x)
  =
  f(r_*) + f'(r_*)x + \frac12 f''(r_*)x^2 + \frac16 f'''(r_*)x^3 + \cdots.
  \label{eq:f_taylor}
\end{equation}
Using $f(r_*)=F(r_*)r_*=0$ from \eqref{eq:fixedpoint_condition}, we obtain
\begin{equation}
  \dot x
  =
  a_1 x + a_2 x^2 + a_3 x^3 + \cdots + \xi(t),
  \label{eq:x_polynomial}
\end{equation}
with coefficients defined by
\begin{equation}
  a_1 := f'(r_*),\qquad
  a_2 := \frac12 f''(r_*),\qquad
  a_3 := \frac16 f'''(r_*),\ \ldots
  \label{eq:a_defs}
\end{equation}

\vspace{10pt}
\emph{(III) Compute $a_1=f'(r_*)$ explicitly (``bare mass/relaxation rate'')\rm{:}}
Since $f(r)=F(r)\,r$, by the product rule
\begin{equation}
  f'(r) = F'(r)\,r + F(r).
  \label{eq:fprime_product}
\end{equation}
At the fixed point $F(r_*)=0$, thus
\begin{equation}
  a_1=f'(r_*) = F'(r_*)\,r_*.
  \label{eq:a1_Fprime}
\end{equation}

Now compute $F'(r)$. From $F(r)=-1+\lambda D(r)^{-1}$,
\begin{equation}
  F'(r) = \lambda\,\frac{d}{dr}\big(D(r)^{-1}\big).
  \label{eq:Fprime_step1}
\end{equation}
Using $D(r)=1+\kappa(r^2-1)$,
\begin{equation}
  D'(r)=2\kappa r.
  \label{eq:Dprime}
\end{equation}
By the chain rule,
\begin{equation}
  \frac{d}{dr}D^{-1} = -(D')D^{-2} = -(2\kappa r)\,D^{-2}.
  \label{eq:Dinv_prime}
\end{equation}
Therefore
\begin{equation}
  F'(r)=\lambda\big(-(2\kappa r)D^{-2}\big)=-2\lambda\kappa r\,D(r)^{-2}.
  \label{eq:Fprime}
\end{equation}
At the fixed point, $D(r_*)=\lambda$ (from $F(r_*)=0$), hence
\begin{equation}
  F'(r_*) = -2\lambda\kappa r_*\,\lambda^{-2}=-\frac{2\kappa r_*}{\lambda}.
  \label{eq:Fprime_at_rstar}
\end{equation}
Plugging into \eqref{eq:a1_Fprime},
\begin{equation}
  a_1 = F'(r_*)\,r_* = -\frac{2\kappa r_*^2}{\lambda}.
  \label{eq:a1_intermediate}
\end{equation}
Using $r_*^2=1+\Delta/\kappa$,
\begin{equation}
  a_1(\lambda,\kappa)= -\frac{2(\kappa+\lambda-1)}{\lambda}.
  \label{eq:a1_final}
\end{equation}
If $\kappa>0$ and $\lambda>1$ then $a_1<0$, meaning the fixed point is linearly stable (IR attractive).

Evaluating the higher-order coefficients at the fixed point $r_*$ yields
\begin{align}
a_2(\lambda,\kappa)
&=
\frac{\sqrt{\kappa(\kappa+\lambda-1)}}{\lambda^2}
\big(4\kappa+\lambda-4\big),
\label{eq:a2_lk}
\\
a_3(\lambda,\kappa)
&=
-\frac{\kappa}{\lambda^3}
\Big(\lambda^2 + 8(\kappa-1)(\kappa+\lambda-1)\Big).
\label{eq:a3_lk}
\end{align}


\section{Appendix C: Fluctuation-induced renormalization within the MSRJD formalism}

We work near the radial fixed point $r_*$ and use the local field
\begin{equation}
  x(t):=r(t)-r_*.
\end{equation}
The stochastic dynamics (Langevin form) is
\begin{equation}
  \dot x(t)=a_1 x(t)+a_2 x(t)^2+a_3 x(t)^3+\xi(t).
  \label{eq:langevin}
\end{equation}

\begin{figure}[t]
\centering
\includegraphics[scale=0.6, trim= 1.0cm 23cm 0cm 0cm]{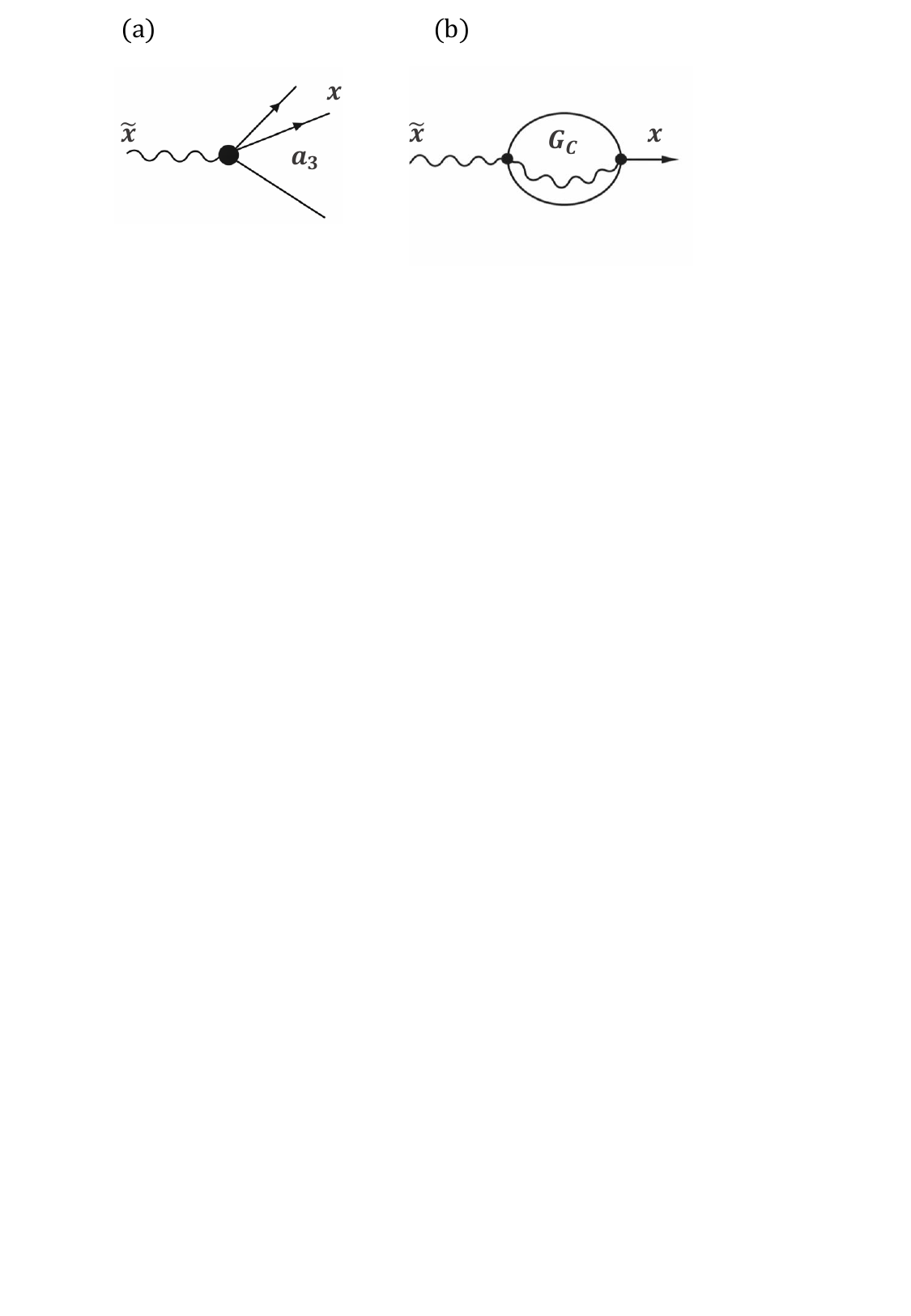}
\caption{One-loop correction to the linear coupling.
(a) Cubic interaction vertex arising from the nonlinear drift term
$\tilde x a_3 x^3$ in the MSRJD action.
(b) Leading one-loop tadpole diagram contributing to the renormalization of
the linear coupling $a_1$.
The loop involves the correlation propagator $G_C$, while the external legs
correspond to the response field $\tilde x$ and the state variable $x$.
Diagrammatic contribution is translated into an analytic expression,
$\delta a_1 \propto -a_3 \int (d\omega/2\pi)\, G_C(\omega)$,
which exhibits a formal infrared divergence as $a_1\to 0$.
}
\label{fig:S1}
\end{figure}

\vspace{10pt}
\emph{(I) MSRJD functional integral\rm{:}}
Equation \eqref{eq:langevin} can be written as a constraint functional:
\begin{equation}
  \mathcal{E}[x,\xi](t):=
  \dot x(t)-a_1x(t)-a_2x(t)^2-a_3x(t)^3-\xi(t)=0.
  \label{eq:constraint}
\end{equation}
Introduce the functional delta to enforce \eqref{eq:constraint}:
\begin{equation}
  \delta\!\big[\mathcal{E}[x,\xi]\big]
  =
  \prod_t \delta\!\Big(\dot x-a_1x-a_2x^2-a_3x^3-\xi\Big).
\end{equation}
For Gaussian noise, the noise weight is
\begin{equation}
  \mathcal{P}[\xi]\propto
  \exp\!\left(
    -\frac{1}{4D}\int dt\,\xi(t)^2
  \right),
  \label{eq:noise_weight}
\end{equation}
since this reproduces $\langle\xi(t)\xi(t')\rangle=2D\delta(t-t')$.

Represent the delta functional by an auxiliary (response) field $\tilde x(t)$:
\begin{equation}
\begin{split}
  &\delta \big[\mathcal{E}\big]
  \propto
  \\
  &\int\mathcal{D}\tilde x
  \exp\!\left(
    \int dt
    \tilde x(t)
    \big[\dot x-a_1x-a_2x^2-a_3x^3-\xi\big]
  \right).
  \label{eq:delta_rep}
\end{split}
\end{equation}

Combine \eqref{eq:noise_weight} and \eqref{eq:delta_rep}:
\begin{equation}
\begin{split}
  &\int\mathcal{D}\xi\;
  \exp\!\left(
    -\frac{1}{4D}\int dt\,\xi^2
    -\int dt\,\tilde x\,\xi
  \right)
  \\
  &\propto
  \exp\!\left(
    -D\int dt\,\tilde x(t)^2
  \right),
  \label{eq:gaussian_int}
\end{split}
\end{equation}
since completing the square gives
\[
-\frac{1}{4D}\xi^2-\tilde x\xi
= -\frac{1}{4D}\big(\xi+2D\tilde x\big)^2 + D\tilde x^2.
\]

After integrating out $\xi$, the MSRJD action becomes
\begin{equation}
  S[x,\tilde x]
  =
  \int dt\;
  \Big[
    \tilde x\big(\dot x-a_1x-a_2x^2-a_3x^3\big)
    -D\tilde x^2
  \Big].
  \label{eq:msrjd_action}
\end{equation}
$\tilde x(t)$ is the response (auxiliary) field.

\vspace{10pt}
\emph{(II) Free theory: bare response and bare correlator\rm{:}}
Split the action into Gaussian (free) + interaction:
\begin{equation}
  S = S_0 + S_{\mathrm{int}},
\end{equation}
with
\begin{equation}
\begin{split}
  &S_0[x,\tilde x]
  =
  \int dt\;
  \Big[
    \tilde x(\dot x-a_1x)
    -D\tilde x^2
  \Big],
  \qquad
  \\
  &S_{\mathrm{int}}[x,\tilde x]
  =
  \int dt\;
  \Big[
    -a_2\tilde x x^2
    -a_3\tilde x x^3
  \Big].
  \label{eq:S0_Sint}
\end{split}
\end{equation}

The bare response is defined as the free-theory contraction
\begin{equation}
  G_0(t-t')
  :=
  \langle x(t)\tilde x(t')\rangle_0.
  \label{eq:G0_def}
\end{equation}
From $S_0$ one finds the linear equation $(\partial_t-a_1)G_0(t-t')=\delta(t-t')$ with retarded boundary condition, giving
\begin{equation}
  G_0(t-t')=\Theta(t-t')e^{a_1(t-t')}.
  \label{eq:G0_time}
\end{equation}
Fourier transform yields
\begin{equation}
  G_0(\omega)=\frac{1}{-i\omega-a_1}.
  \label{eq:G0_freq}
\end{equation}

Define the free correlator
\begin{equation}
  C_0(\omega):=\langle x(\omega)x(-\omega)\rangle_0.
  \label{eq:C0_def}
\end{equation}
Using $x(\omega)=G_0(\omega)\xi(\omega)$ in the equivalent Langevin picture and $\langle\xi(\omega)\xi(\omega')\rangle=2D(2\pi)\delta(\omega+\omega')$, one obtains
\begin{equation}
  C_0(\omega)=2D\,|G_0(\omega)|^2
  =
  \frac{2D}{\omega^2+a_1^2}.
  \label{eq:C0}
\end{equation}

\vspace{10pt}
\emph{(III) Vertices (interaction terms)\rm{:}}
From $S_{\mathrm{int}}$ we read off the vertices:
\begin{equation}
  \text{2-vertex}\, (V_2): -a_2\,\tilde x\,x^2,
  \qquad
  \text{3-vertex}\, (V_3): -a_3\,\tilde x\,x^3.
  \label{eq:vertices}
\end{equation}
Every vertex carries one $\tilde x$ and two/three $x$ legs, respectively.

\vspace{10pt}
\emph{(IV) Wilsonian step: split into slow/fast modes and integrate out fast shell\rm{:}}
Introduce a UV frequency cutoff $\Lambda$ and decompose fields:
\begin{equation}
  x = x_< + x_>,\qquad \tilde x=\tilde x_<+\tilde x_>,
\end{equation}
where $ x_>(\omega),\tilde x_>(\omega)$ live in the shell  $ \Lambda/b<|\omega|<\Lambda$,
\begin{equation}
  b=e^{d\ell}\approx 1+d\ell,
  \label{eq:shell}
\end{equation}
and $x_<,\tilde x_<$ contain modes $|\omega|<\Lambda/b$.

The effective action for slow modes is defined by
\begin{equation}
\begin{split}
  &e^{-S_{\mathrm{eff}}[x_<,\tilde x_<]}
  \\
  &:=
  \int \mathcal{D}x_>\mathcal{D}\tilde x_>\;
  e^{-S_0[x_<+x_>,\tilde x_<+\tilde x_>]-S_{\mathrm{int}}[x_<+x_>,\tilde x_<+\tilde x_>]}.
  \label{eq:Seff_def}
\end{split}
\end{equation}
Expand perturbatively in $S_{\mathrm{int}}$:
\begin{equation}
\begin{split}
  &S_{\mathrm{eff}} = S_0[x_<,\tilde x_<] - \ln\Big\langle e^{-S_{\mathrm{int}}}\Big\rangle_>
  \\
  &=
  S_0[x_<,\tilde x_<]
  + \Big\langle S_{\mathrm{int}}\Big\rangle_>
  -\frac{1}{2}\Big(
    \big\langle S_{\mathrm{int}}^2\big\rangle_>-\big\langle S_{\mathrm{int}}\big\rangle_>^2
  \Big)+\cdots,
  \label{eq:cumulant}
\end{split}
\end{equation}
where $\langle\cdots\rangle_>$ denotes averaging with the \emph{fast} Gaussian measure $S_0[x_>,\tilde x_>]$.

\vspace{10pt}
\emph{(V) 1-loop renormalization of $a_1$ from the $a_3$ tadpole\rm{:}}
We focus on the leading contribution from the cubic vertex
\begin{equation}
  S_3 := \int dt\,\big(-a_3\tilde x\,x^3\big).
  \label{eq:S3}
\end{equation}
Insert the slow/fast split into $S_3$:
\begin{equation}
  -a_3\int dt\,
  (\tilde x_<+\tilde x_>)\,(x_<+x_>)^3.
  \label{eq:S3_split}
\end{equation}
Expand the cube:
\begin{equation}
  (x_<+x_>)^3 = x_<^3 + 3x_<^2x_> + 3x_<x_>^2 + x_>^3.
  \label{eq:cube_expand}
\end{equation}
We now average over the fast fields. Because $x_>$ is Gaussian with zero mean,
\begin{equation}
  \langle x_>(t)\rangle_> = 0,
  \qquad
  \langle x_>^3\rangle_>=0.
\end{equation}
Thus in $\langle S_3\rangle_>$, only terms involving $x_>^2$ survive:
\begin{equation}
\begin{split}
  &\big\langle S_3 \big\rangle_>
  =
  \\
  &-a_3\int dt\,
  \Big[
    \tilde x_<\,\big\langle 3x_<x_>^2\big\rangle_>
    \ +\ \tilde x_>\,\big\langle 3x_<x_>^2\big\rangle_>
    \ +\ \cdots
  \Big],
  \label{eq:avg_S3_step}
\end{split}
\end{equation}
where $\cdots$ denotes terms that vanish by odd moments or do not contribute to the slow effective action after integrating out $\tilde x_>$.

Keeping the slow operator structure $\tilde x_< x_<$,
\begin{equation}
  \big\langle S_3 \big\rangle_>
  \supset
  -3a_3\int dt\;
  \tilde x_<(t)\,x_<(t)\;
  \underbrace{\big\langle x_>(t)^2\big\rangle_>}_{=: \langle x^2\rangle_>}.
  \label{eq:mass_shift_operator}
\end{equation}

In the slow action, the linear term is
\begin{equation}
  \int dt\,\tilde x_<\big(\dot x_<-a_1 x_<\big).
\end{equation}
A correction of the form
\(
-\int dt\,(\delta a_1)\,\tilde x_<x_<
\)
is equivalent to shifting $a_1\to a_1+\delta a_1$.
Comparing with \eqref{eq:mass_shift_operator}, we identify
\begin{equation}
  \delta a_1 = 3a_3\,\langle x^2\rangle_>.
  \label{eq:delta_a1_basic}
\end{equation}
(The overall sign can flip depending on the $\tilde x$ convention; the \emph{structure} and the factor $3$ from combinatorics are robust.)

The fast-mode equal-time variance is
\begin{equation}
  \langle x^2\rangle_>
  :=
  \langle x_>(t)x_>(t)\rangle_>
  =
  \int_{\Lambda/b<|\omega|<\Lambda}\frac{d\omega}{2\pi}\;C_0(\omega),
  \label{eq:x2_shell_def}
\end{equation}
where $C_0(\omega)$ is the \emph{bare} correlator \eqref{eq:C0}:
\begin{equation}
  C_0(\omega)=\frac{2D}{\omega^2+a_1^2}.
\end{equation}
Therefore
\begin{equation}
  \langle x^2\rangle_>
  =
  \int_{\Lambda/b<|\omega|<\Lambda}\frac{d\omega}{2\pi}\;
  \frac{2D}{\omega^2+a_1^2}.
  \label{eq:x2_shell_integral}
\end{equation}
Using symmetry in $\omega$,
\begin{equation}
  \int_{\Lambda/b<|\omega|<\Lambda} d\omega\,(\cdots)
  = 2\int_{\Lambda/b}^{\Lambda} d\omega\,(\cdots),
\end{equation}
we obtain
\begin{equation}
  \langle x^2\rangle_>
  =
  \frac{2D}{2\pi}\cdot 2
  \int_{\Lambda/b}^{\Lambda}\frac{d\omega}{\omega^2+a_1^2}
  =
  \frac{2D}{\pi}
  \int_{\Lambda/b}^{\Lambda}\frac{d\omega}{\omega^2+a_1^2}.
  \label{eq:x2_shell_simplified}
\end{equation}

Define
\begin{equation}
  I:=\int_{\Lambda/b}^{\Lambda}\frac{d\omega}{\omega^2+a_1^2}.
\end{equation}
For a thin shell, approximate the integrand by its value near $\omega=\Lambda$:
\begin{equation}
  I \approx \frac{1}{\Lambda^2+a_1^2}\int_{\Lambda/b}^{\Lambda}d\omega
  =
  \frac{\Lambda-\Lambda/b}{\Lambda^2+a_1^2}
  =
  \frac{\Lambda(1-1/b)}{\Lambda^2+a_1^2}.
  \label{eq:I_thinshell1}
\end{equation}
With $b=e^{d\ell}$, we have $1/b=e^{-d\ell}\approx 1-d\ell$, hence
\begin{equation}
  1-\frac{1}{b}\approx d\ell,
\end{equation}
and therefore
\begin{equation}
  I \approx \frac{\Lambda}{\Lambda^2+a_1^2}\,d\ell.
  \label{eq:I_thinshell2}
\end{equation}
Substituting into \eqref{eq:x2_shell_simplified} gives
\begin{equation}
  \langle x^2\rangle_>
  \approx
  \frac{2D}{\pi}\,
  \frac{\Lambda}{\Lambda^2+a_1^2}\,d\ell.
  \label{eq:x2_thinshell}
\end{equation}

From \eqref{eq:delta_a1_basic} and \eqref{eq:x2_thinshell},
\begin{equation}
  \delta a_1
  \approx
  3a_3\cdot
  \frac{2D}{\pi}\,
  \frac{\Lambda}{\Lambda^2+a_1^2}\,d\ell.
\end{equation}
Hence the 1-loop contribution to the beta function is
\begin{equation}
  \beta_{a_1}^{(1)}:=\frac{da_1}{d\ell}\Big|_{\mathrm{1-loop}}
  \approx
  3a_3\cdot
  \frac{2D}{\pi}\,
  \frac{\Lambda}{\Lambda^2+a_1^2}.
  \label{eq:beta_a1_1loop}
\end{equation}

In 0+1D, the shell integral typically yields a cutoff-dependent factor
$\sim \Lambda/(\Lambda^2+a_1^2)$ rather than a universal logarithm.
The Wilsonian step is well-defined, but numerical coefficients can be scheme/cutoff sensitive.

\vspace{10pt}
\emph{(VI) Tree-level rescaling contribution\rm{:}}
If we rescale time as $t'=bt$ with $b=e^{d\ell}$, then $\partial_t = b\,\partial_{t'}$.
Keeping the kinetic structure $\tilde x\,\dot x$ fixed without field rescaling implies the drift coefficient scales as
\begin{equation}
  a_1' = b\,a_1,
\end{equation}
so
\begin{equation}
  \beta_{a_1}^{(\mathrm{tree})}=\frac{da_1}{d\ell}\Big|_{\mathrm{tree}}=a_1.
  \label{eq:beta_a1_tree}
\end{equation}
Thus a minimal combined flow reads
\begin{equation}
  \beta_{a_1}
  =
  a_1
  +
  3a_3\cdot
  \frac{2D}{\pi}\,
  \frac{\Lambda}{\Lambda^2+a_1^2}
  +\cdots,
  \label{eq:beta_a1_total}
\end{equation}
where $\cdots$ denotes additional 1-loop terms (e.g.\ from $a_2$) and possible $D$ renormalization.

\vspace{10pt}
\emph{(VII) Derivation of the $\Delta$-beta function near the critical manifold\rm{:}}
We parameterize
\begin{equation}
\lambda = 1+\Delta,
\qquad
\kappa = 1+\varepsilon,
\qquad
|\Delta|,|\varepsilon|\ll 1 .
\end{equation}

The condition $\lambda=1$ defines the critical shell location in the reduced
description. The homeostatic curvature $\kappa$ remains a physical structural
parameter; in the following we evaluate nonuniversal prefactors at a fixed
reference value $\kappa=1$ in order to present the resulting flow in a canonical
normal form.

At $\kappa=1$, the cubic coefficient admits the expansion
\begin{equation}
a_3(\kappa=1)
=
-1 + \Delta - \Delta^2 + \mathcal O(\Delta^3).
\end{equation}

We then define the dimensionless deviation from criticality as
\begin{equation}
\tilde\Delta := g_{\mathrm{eff}}\,\Delta,
\qquad
g_{\mathrm{eff}}
=
\frac{6D}{\pi}\frac{\Lambda}{\Lambda^2+a_1^2},
\end{equation}
where $g_{\mathrm{eff}}$ collects nonuniversal factors arising from the
Wilsonian shell integration.

With this definition, the beta function assumes the canonical normal form
\begin{equation}
\beta_{\tilde\Delta}
=
\frac{d\tilde\Delta}{d\ell}
=
\tilde\Delta
-
\tilde\Delta^{\,2}
+
\mathcal O(\tilde\Delta^{\,3}),
\end{equation}
demonstrating that the critical manifold remains infrared attractive and
marginally stable at one-loop order.

\section{Appendix D: Translating the cubic-vertex RG Flow to $\beta_\kappa$ and extracting $\alpha_i$}

We expand near the scheme/reference point
\begin{equation}
(\lambda,\kappa)=(1,1),
\quad
\lambda=1+\Delta,\quad \kappa=1+\varepsilon,
\quad |\Delta|,|\varepsilon|\ll 1.
\end{equation}
We use the (scheme) condition that defines the renormalized cubic vertex at the
reference point:
\begin{equation}
a_3^{R}(\Delta=0,\kappa=1)\equiv -1
\quad \text{for all RG scales }\ell.
\label{eq:G2}
\end{equation}

The one-loop renormalization of $a_3$ comes from the connected cumulant
$\frac12\langle S_{\text{int}}^2\rangle_c$ (the mean $\langle S_{\text{int}}\rangle$
only generates tadpoles that renormalize $a_1$ or induce a shift of the mean).
At one loop, the 1PI correction to $\Gamma^{(1,3)}$ receives contributions
from the following topologies:

\vspace{10pt}
\emph{(I) $a_2 a_3$ mixed vertex correction\rm{:}}
One $V_2$ and one $V_3$ vertex can combine to generate an effective
$\tilde x x^3$ operator after contracting fast legs.
The corresponding one-loop contribution has the schematic structure
\begin{equation}
\delta a_3\big|_{a_2a_3}
=
c_{23}^{(1)}
\,a_2 a_3\,
I_{23}^{(1)},
\label{eq:da3_a2a3}
\end{equation}
where the shell integral at the static renormalization point is
\begin{equation}
I_{23}^{(1)}
:=
\int_{\text{shell}}\frac{d\omega}{2\pi}\,
G_0(\omega)\,C_0(\omega),
\label{eq:I23_def}
\end{equation}
and $c_{23}^{(1)}$ is the combinatorial factor
(number of distinct contractions producing $\tilde x x^3$).

Using
\begin{equation}
G_0(\omega)C_0(\omega)
=
\frac{2D}{(-i\omega-a_1)(\omega^2+a_1^2)},
\end{equation}
the shell integral yields (thin-shell approximation)
\begin{equation}
I_{23}^{(1)}
\simeq
\frac{2D}{\pi}\,
\frac{(-a_1)\Lambda}{(\Lambda^2+a_1^2)^2}\,d\ell.
\quad
\label{eq:I23_eval}
\end{equation}
Odd imaginary part cancels in symmetric shell.

\vspace{10pt}
\emph{(II) $a_2^2$ induced cubic vertex via shift removal\rm{:}}
The quadratic vertex $V_2$ generically generates a nonzero mean (a ``tadpole''
in the equation of motion). After removing the mean by a field shift
$x\to x+\bar x$ (so that $\langle x\rangle=0$), an effective cubic operator
is induced. This effect is captured by
\begin{equation}
\delta a_3\big|_{\text{shift}}
=
c_{22}^{(1)}\,a_2^2\,I_{22}^{(1)},
\label{eq:da3_shift}
\end{equation}
with a shell integral of the form
\begin{equation}
I_{22}^{(1)} := \int_{\text{shell}}\frac{d\omega}{2\pi}\,C_0(\omega)
\simeq
\frac{2D}{\pi}\frac{\Lambda}{\Lambda^2+a_1^2}\,d\ell.
\label{eq:I22_eval}
\end{equation}
(Here $c_{22}^{(1)}$ depends on the precise convention for shift subtraction,
but becomes unique once the ``$\langle x\rangle=0$'' renormalization condition
is imposed.)

\vspace{10pt}
\emph{(III) $a_3^2$ one-loop vertex correction\rm{:}}
Two $V_3$ vertices also contribute to $\Gamma^{(1,3)}$ at one loop.
The schematic form is
\begin{equation}
\delta a_3\big|_{a_3^2}
=
c_{33}^{(1)}\,a_3^2\,I_{33}^{(1)},
\label{eq:da3_a3a3}
\end{equation}
where the basic building block integral at zero external frequency is
\begin{equation}
I_{33}^{(1)}
:=
\int_{\text{shell}}\frac{d\omega}{2\pi}\,
G_0(\omega)\,C_0(\omega)
\;\;\; \text{(same scaling as $I_{23}^{(1)}$)}.
\label{eq:I33_def}
\end{equation}
In thin-shell form,
\begin{equation}
I_{33}^{(1)}\simeq I_{23}^{(1)}
\simeq
\frac{2D}{\pi}\,
\frac{(-a_1)\Lambda}{(\Lambda^2+a_1^2)^2}\,d\ell.
\end{equation}

\vspace{10pt}
\emph{(IV) One-loop beta function for $a_3$\rm{:}}
Collecting one-loop contributions and including the tree-level scaling
($a_3$ has the same canonical time dimension as $a_1$ in our convention),
\begin{equation}
\begin{split}
&\beta_{a_3}
\equiv
\frac{da_3}{d\ell}
=
\\
&a_3
+
c_{22}^{(1)}a_2^2 I_{22}^{(1)}
+
c_{23}^{(1)}a_2 a_3 I_{23}^{(1)}
+
c_{33}^{(1)}a_3^2 I_{33}^{(1)}
+
\mathcal O(\text{2-loop}).
\label{eq:beta_a3_1loop}
\end{split}
\end{equation}
All coefficients become unique once the renormalization conditions and
vertex normalization plus the ``$\langle x\rangle=0$'' shift-fixing
condition are imposed.

\vspace{10pt}
\emph{(V) Translation to $\beta_\kappa$ and the coefficients $\alpha_i$\rm{:}}
We expand near the reference point
\begin{equation}
\lambda=1+\Delta,\qquad \kappa=1+\varepsilon,
\qquad |\Delta|,|\varepsilon|\ll 1,
\end{equation}
and define the dimensionless deviation $\tilde\Delta$ as in the previous section
so that
\begin{equation}
\beta_{\tilde\Delta}
=
\tilde\Delta-\tilde\Delta^2+\mathcal O(\tilde\Delta^3).
\end{equation}

Since $a_3=a_3(\Delta,\kappa)$, the chain rule gives
\begin{equation}
\beta_{a_3}
=
\frac{\partial a_3}{\partial \Delta}\,\beta_\Delta
+
\frac{\partial a_3}{\partial \kappa}\,\beta_\kappa.
\label{eq:chain_a3}
\end{equation}
At the scheme point $(\Delta,\kappa)=(0,1)$, condition (S87) enforces
$a_3^R\equiv -1$ for all $\ell$, hence
\begin{equation}
\beta_{a_3}\big|_{(0,1)}=0.
\label{eq:beta_a3_zero}
\end{equation}
Expanding Eq.~\eqref{eq:chain_a3} to first nontrivial orders in
$(\varepsilon,\tilde\Delta)$ yields the normal form
\begin{equation}
\beta_\kappa
=
\alpha_1(\kappa-1)
+
\alpha_2\,\tilde\Delta
+
\alpha_3\,\tilde\Delta^2
+\cdots,
\end{equation}
where the coefficients $\alpha_i$ are obtained by matching the expansion of
$\beta_{a_3}$.

Concretely, at one loop,
$\alpha_1,\alpha_2,\alpha_3$ are linear combinations of
$( c_{22}^{(1)}a_2^2 I_{22}^{(1)},\;
c_{23}^{(1)}a_2 a_3 I_{23}^{(1)},\;
c_{33}^{(1)}a_3^2 I_{33}^{(1)} ) $
evaluated near $(\Delta,\kappa)=(0,1)$, plus the scheme-fixed canonical scaling.

Once a specific shift-fixing convention (e.g.\ enforcing $\langle x\rangle=0$ at each RG step)
is implemented, the combinatorial coefficients
$c_{22}^{(1)},c_{23}^{(1)},c_{33}^{(1)}$ become uniquely determined,
and the integrals $I^{(1)}$ and $I^{(2)}$ can be evaluated in closed form
(or thin-shell approximations) to yield explicit numerical values for
$\alpha_1,\alpha_2,\alpha_3$.

\vspace{10pt}
\emph{(VI) Explicit one-loop computation of $\alpha_i$\rm{:}}
At one loop, the cubic vertex receives corrections from
mixed $(a_2 a_3)$ and self $(a_3^2)$ diagrams.
Collecting all contractions consistent with causality,
the correction takes the form
\begin{equation}
\delta a_3
=
\bigl(6a_2a_3+9a_3^2\bigr)\,I_C .
\end{equation}

Including tree-level scaling, we obtain
\begin{equation}
\beta_{a_3}
=
\frac{da_3}{d\ell}
=
a_3
+
\bigl(6a_2a_3+9a_3^2\bigr)\,
\frac{2D}{\pi}
\frac{\Lambda}{\Lambda^2+a_1^2}
+
\mathcal O(\text{2-loop}) .
\end{equation}

The physical flow is defined by subtracting the reference-point contribution,
\begin{equation}
\beta_{a_3}^{\mathrm{phys}}(\Delta,\kappa)
=
\beta_{a_3}(\Delta,\kappa)
-
\beta_{a_3}(0,1),
\end{equation}
which enforces condition (S87).

Near the reference point in Eq. (S98),
the expansion of $a_3(\Delta,\kappa)$ yields
\begin{equation}
\left.
\frac{\partial a_3}{\partial \kappa}
\right|_{(0,1)}
=-9,
\qquad
\left.
\frac{\partial a_3}{\partial \Delta}
\right|_{(0,1)}
=-5 .
\end{equation}

Using Eq. (S99), we obtain the normal-form flow (S102), with one-loop coefficients
\begin{align}
\alpha_1 &= 1 - \frac{16D}{\pi}
\frac{\Lambda}{\Lambda^2+a_1^2}, \\
\alpha_2 &= -\frac{20}{3}\,\frac{1}{g}
\frac{2D}{\pi}
\frac{\Lambda}{\Lambda^2+a_1^2}, \\
\alpha_3 &= \text{scheme-dependent at one loop}.
\end{align}

All coefficients are fully determined by the microscopic parameters
$(D,\Lambda,a_1)$.

\section{Appendix E: Stationary variance of the linearized radial dynamics}

In the vicinity of the stable shell, the radial fluctuations
$\delta r(t)=r(t)-r_\ast$ obey a linear stochastic differential equation of
Ornstein--Uhlenbeck type,
\begin{equation}
\delta \dot r(t) = a\,\delta r(t) + \xi(t),
\label{eq:ou_app}
\end{equation}
where $a=a(\lambda,\kappa)<0$ is the linear relaxation rate and $\xi(t)$ denotes
Gaussian white noise.

Equation~\eqref{eq:ou_app} is solved most transparently using an integrating
factor.
Multiplying both sides by $e^{-at}$ yields
\begin{equation}
\frac{d}{dt}\!\left(e^{-at}\,\delta r(t)\right)
=
e^{-at}\,\xi(t).
\end{equation}
Integrating from $t'=-\infty$ to $t$ and imposing the stability condition
$a<0$, which ensures that contributions from the remote past decay, we obtain
\begin{equation}
e^{-at}\,\delta r(t)
=
\int_{-\infty}^{t} dt'\,
e^{-a t'}\,\xi(t').
\end{equation}
Rearranging gives the causal solution
\begin{equation}
\delta r(t)
=
\int_{-\infty}^{t} dt'\,
e^{a(t-t')}\,\xi(t').
\label{eq:ou_solution_app}
\end{equation}

This expression can be written compactly as a convolution with the retarded
Green's function,
\begin{equation}
G_R(t-t') = \Theta(t-t')\,e^{a(t-t')},
\end{equation}
so that $\delta r = G_R * \xi$, consistent with the MSRJD response formalism.

\emph{(I) Stationary variance\rm{:}}
Using Eq.~\eqref{eq:ou_solution_app} and the noise correlator, the stationary variance is computed as
\begin{align}
\mathrm{Var}[\delta r]
&=
\langle (\delta r(t))^2 \rangle
\nonumber\\
&=
\int_{-\infty}^{t} dt_1
\int_{-\infty}^{t} dt_2\,
e^{a(2t-t_1-t_2)}
\langle \xi(t_1)\xi(t_2) \rangle
\nonumber\\
&=
2D_{\mathrm{tot}}
\int_{-\infty}^{t} dt_1\,
e^{2a(t-t_1)}
\nonumber\\
&=
2D_{\mathrm{tot}}
\int_{0}^{\infty} ds\, e^{2as}
\nonumber\\
&=
\frac{D_{\mathrm{tot}}}{-a}.
\label{eq:variance_app}
\end{align}
The variance is finite provided $a<0$, confirming that the shell is linearly
stable and that fluctuations remain bounded.

\section{Appendix F: OU linearization and MSRJD closure for learning-induced flow}

\emph{(I) Ornstein-Uhlenbeck (OU) linearization around the stable shell\rm{:}}
Let
\begin{equation}
\label{eq:delta_r_def}
r(t)=r_*+\delta r(t).
\end{equation}
Taylor-expand the drift:
\begin{equation}
\label{eq:taylor_f}
f(r_*+\delta r)=f(r_*)+\partial_r f(r)\big|_{r_*}\,\delta r + O(\delta r^2).
\end{equation}
Since $f(r_*)=0$, dropping $O(\delta r^2)$ yields the OU approximation:
\begin{equation}
\label{eq:ou_sde}
\delta\dot r(t)=a\,\delta r(t)+h(t)+\xi(t),
\qquad
a:=\partial_r f(r;\lambda,\kappa)\big|_{r=r_*}.
\end{equation}

One can compute $a$ explicitly. Using the identity
$1+\kappa(r_*^2-1)=\lambda$, the standard algebra gives
\begin{equation}
\label{eq:a_exact}
a=-\frac{2\kappa r_*^2}{\lambda}.
\end{equation}
Thus for $\kappa>0$ the shell is linearly stable ($a<0$), and the OU relaxation time is
$\tau_r\sim |a|^{-1}$.

\vspace{10pt}
\emph{(II) Choice of teaching signal and reduction to $\langle r^2-1\rangle$\rm{:}}
We choose a minimal homeostatic teaching/source term that penalizes deviation from the
target shell $r\simeq 1$:
\begin{equation}
\label{eq:h_teach_choice}
h(t)=h_{\rm teach}(t):=-\epsilon\,(r(t)^2-1),
\qquad \epsilon>0,
\end{equation}
treated as slowly varying on the fast OU time scale.

From the radial drift,
\begin{equation}
\label{eq:dlambda_f_exact}
\partial_\lambda f(r;\lambda,\kappa)=\frac{r}{1+\kappa(r^2-1)}.
\end{equation}
Expand around $r=r_*+\delta r$:
\begin{equation}
\begin{split}
\label{eq:dlambda_expand}
&\partial_\lambda f(r)\approx B_* + B_1\,\delta r,
\qquad
B_*:=\partial_\lambda f(r_*)=\frac{r_*}{\lambda},
\qquad
\\
&B_1:=\partial_r\partial_\lambda f(r)\big|_{r_*}.
\end{split}
\end{equation}
The leading nontrivial contribution to $\langle \tilde r\,\partial_\lambda f\rangle$
arises from the $\delta r$-dependent part (the constant part typically does not generate a
nonzero stationary average under causal/It\^{o} conventions without a drive):
\begin{equation}
\label{eq:key_reduction1}
\langle \tilde r\,\partial_\lambda f\rangle
\approx
B_1\,\langle \tilde r\,\delta r\rangle.
\end{equation}

In OU, the mean fluctuation is given by linear response:
\begin{equation}
\label{eq:delta_r_mean}
\langle \delta r(t)\rangle
=
\int dt'\,R(t,t')\,\langle h(t')\rangle.
\end{equation}
For slow $h$ we apply the DC gain:
\begin{equation}
\label{eq:delta_r_mean_DC}
\langle \delta r(t)\rangle\approx -\frac{1}{a}\,\langle h(t)\rangle.
\end{equation}
Heuristically (and consistently with quadratic MSRJD structure), the same retarded kernel
controls the mixed response insertion, so the nontrivial part of $\langle \tilde r\,\delta r\rangle$
scales with the same DC factor. We therefore parameterize
\begin{equation}
\label{eq:rtildedr_scaling}
\langle \tilde r\,\delta r\rangle
=
C_{\rm OU}\,\Big(-\frac{1}{a}\Big)\,\langle h\rangle,
\qquad C_{\rm OU}>0,
\end{equation}
where $C_{\rm OU}$ absorbs convention-dependent equal-time regularization (It\^{o}/Stratonovich)
and normalization constants. This is sufficient for determining the induced \emph{sign} and
leading scaling of the slow drift.

Combining \eqref{eq:key_reduction1} and \eqref{eq:rtildedr_scaling},
\begin{equation}
\label{eq:key_reduction2}
\langle \tilde r\,\partial_\lambda f\rangle
\approx
B_1\,C_{\rm OU}\,\Big(-\frac{1}{a}\Big)\,\langle h\rangle.
\end{equation}

Using the teaching choice \eqref{eq:h_teach_choice},
\begin{equation}
\label{eq:h_mean}
\langle h\rangle
=
-\epsilon\,\langle r^2-1\rangle.
\end{equation}
Under time-scale separation and small noise (fast measure concentrated near $r_*$),
\begin{equation}
\label{eq:r2_close}
\langle r^2-1\rangle
\approx
r_*^2-1.
\end{equation}
Substituting \eqref{eq:h_mean} into \eqref{eq:key_reduction2} yields the desired reduction:
\begin{equation}
\label{eq:target_derived}
\langle \tilde r\,\partial_\lambda f\rangle
\;\propto\;
-\langle r^2-1\rangle,
\end{equation}
where the proportionality constant is
\begin{equation}
K_\lambda
:=
B_1\,C_{\rm OU}\,\Big(-\frac{1}{a}\Big)\,\epsilon,
\end{equation}
and can be restricted to the regime where $K_\lambda>0$ (stable shell, weak teaching, and
our causal convention).

\vspace{10pt}
\emph{(III) Induced slow flow for $\Delta=\lambda-1$\rm{:}}
From the learning drift with $\theta=\lambda$ and
\eqref{eq:target_derived},
\begin{equation}
\langle \dot\lambda\rangle
=
\eta\,\langle \tilde r\,\partial_\lambda f\rangle
\approx
-\eta\,K_\lambda\,\langle r^2-1\rangle.
\end{equation}
Using the quasi-static closure \eqref{eq:r2_close},
\begin{equation}
\langle r^2-1\rangle \approx r_*^2-1=\frac{\Delta}{\kappa},
\end{equation}
we obtain the beta-function form
\begin{equation}
\label{eq:beta_delta}
\beta_\Delta(\Delta,\kappa)
:=
\langle \dot\Delta\rangle
=
\langle \dot\lambda\rangle
\approx
-c_\Delta\,\eta\,\frac{\Delta}{\kappa},
\qquad
c_\Delta:=K_\lambda>0,
\end{equation}
implying the attractive sign structure (for $\kappa>0$):
\begin{equation}
\Delta>0\Rightarrow \dot\Delta<0,
\qquad
\Delta<0\Rightarrow \dot\Delta>0,
\end{equation}
so $\Delta=0$ is an attractive fixed point of the slow learning flow.


\begin{thebibliography}{99}


\bibitem{1} J. J. Hopfield, ``Neural networks and physical systems with emergent collective computational abilities,'' Proc. Natl. Acad. Sci. USA \textbf{79}, 2554-2558 (1982).
\bibitem{2} D. J. Amit and H. Gutfreund, ``Storing infinite numbers of patterns in a spin-glass model of neural networks,'' Phys. Rev. Lett. \textbf{55}, 1530-1533 (1985).
\bibitem{3} G. E. Hinton and D. C. Plaut, ``Using fast weights to deblur old memories,'' Proc. 9th Annu. Conf. Cognitive Science Society, 177-186 (1987).
\bibitem{4} J. Schmidhuber, ``Learning to control fast-weight memories: An alternative to dynamic recurrent networks,'' Neural Comput. \textbf{4}, 131-139 (1992).
\bibitem{5} J. Ba, G. E. Hinton, V. Mnih, J. Leibo, and C. Ionescu, ``Using fast weights to attend to the recent past,'' arXiv:1610.06258 (2016).
\bibitem{6} I. Schlag, T. Irie, and J. Schmidhuber, ``Linear transformers are secretly fast weight programmers,'' arXiv:2102.11174 (2021).
\bibitem{7} A. Katharopoulos, A. Vyas, N. Pappas, and F. Fleuret, ``Transformers are RNNs: Fast autoregressive transformers with linear attention,'' arXiv:2006.16236 (2020).

\bibitem{8} G. M. Edelman, \emph{Neural Darwinism: The Theory of Neuronal Group Selection} (Basic Books, New York, 1989).
\bibitem{9} G. Tononi, O. Sporns, and G. M. Edelman, ``A measure for brain complexity: Relating functional segregation and integration in the nervous system,''
Proc. Natl. Acad. Sci. USA \textbf{91}, 5033-5037 (1994).
\bibitem{10} G. G. Turrigiano and S. B. Nelson, ``Hebb and homeostasis in neuronal plasticity,'' Curr. Opin. Neurobiol. \textbf{10}, 358-364 (2000).
\bibitem{11} L. F. Abbott and W. G. Regehr, ``Synaptic computation,'' Nature \textbf{431}, 796-803 (2004).
\bibitem{12} E. Oja, ``A simplified neuron model as a principal component analyzer,'' J. Math. Biol. \textbf{15}, 267-273 (1982).

\bibitem{13} B. G. Chae, ``Recursive dynamics in fast-weights homeostatic reentry networks: Toward reflective intelligence,'' arXiv:2511.06798 (2025).
\bibitem{14} B. G. Chae, ``Continuous-time homeostatic dynamics for reentrant inference models,'' arXiv:2512.05158 (2025).
\bibitem{15} B. G. Chae, ``Renormalization-group geometry of homeostatically regulated reentry networks,'' arXiv:2512.19086 (2025).

\bibitem{16} K. Funahashi and Y. Nakamura, ``Approximation of dynamical systems by continuous time recurrent neural networks,''
Neural Networks \textbf{6}, 801-806 (1993).
\bibitem{17} R. D. Beer, ``On the dynamics of small continuous-time recurrent neural networks,'' Adaptive Behavior \textbf{3}, 469-509 (1995).
\bibitem{18} R. Hasani, M. Lechner, A. Amini, D. Rus, and R. Grosu, ``Liquid time-constant networks,'' Proc. AAAI Conf. Artif. Intell. \textbf{35}, 7657-7666 (2021).

\bibitem{19} P. Cannon and J. Miller, ``Synaptic and intrinsic homeostasis cooperate to optimize single neuron response properties and tune integrator circuits,''
J. Neurophysiol. \textbf{116}, 2004-2022 (2016).
\bibitem{20} N. Niemeyer, J. H. Schleimer, and S. Schreiber, ``Biophysical models of intrinsic homeostasis: Firing rates and beyond,''
Curr. Opin. Neurobiol. \textbf{70}, 81-88 (2021).


\bibitem{21} K. G. Wilson and J. Kogut, ``The renormalization group and the $\epsilon$ expansion,'' Phys. Rep. \textbf{12}, 75-199 (1974).
\bibitem{22} J. Cardy, \emph{Scaling and Renormalization in Statistical Physics} (Cambridge University Press, Cambridge, 1996).
\bibitem{23} U. C. T\"auber, \emph{Critical Dynamics} (Cambridge University Press, Cambridge, 2014).

\bibitem{24} P. C. Martin, E. D. Siggia, and H. A. Rose, ``Statistical dynamics of classical systems,'' Phys. Rev. A \textbf{8}, 423-437 (1973).
\bibitem{25} H. K. Janssen, ``On a Lagrangian for classical field dynamics and renormalization group calculations of dynamical critical properties,''
Z. Phys. B \textbf{23}, 377-380 (1976).
\bibitem{26} C. De Dominicis, ``Techniques de renormalisation de la th\'eorie des champs et dynamique des ph\'enom\`enes critiques,''
J. Phys. Colloq. \textbf{37}, 247-253 (1976).

\bibitem{27} J. Zinn-Justin, \emph{Quantum Field Theory and Critical Phenomena} (Oxford University Press, Oxford, 2002).

\bibitem{28} U. Seifert, ``Stochastic thermodynamics, fluctuation theorems and molecular machines,'' Rep. Prog. Phys. \textbf{75}, 126001 (2012).
\bibitem{29} J. M. Horowitz and M. Esposito, ``Thermodynamics with continuous information flow,'' Phys. Rev. X \textbf{4}, 031015 (2014).

\bibitem{30} P. Bak, C. Tang, and K. Wiesenfeld, ``Self-organized criticality: An explanation of $1/f$ noise,'' Phys. Rev. Lett. \textbf{59}, 381-384 (1987).
\bibitem{31} P. Bak, \emph{How Nature Works: The Science of Self-Organized Criticality} (Springer, New York, 1996).
\bibitem{32} H. J. Jensen, \emph{Self-Organized Criticality: Emergent Complex Behavior in Physical and Biological Systems}
(Cambridge University Press, Cambridge, 1998).

\bibitem{33} R. Dickman, M. A. Mu\~noz, A. Vespignani, and S. Zapperi, ``Paths to self-organized criticality,'' Braz. J. Phys. \textbf{30}, 27-41 (2000).

\bibitem{34} S. Bornholdt and T. Rohlf, ``Topological evolution of dynamical networks: Global criticality from local dynamics,''
Phys. Rev. Lett. \textbf{84}, 6114-6116 (2000).
\bibitem{35} A. Levina, J. M. Herrmann, and T. Geisel, ``Phase transitions towards criticality in a neural system with adaptive interactions,''
Phys. Rev. Lett. \textbf{102}, 118110 (2009).
\bibitem{36} D. R. Chialvo, ``Emergent complex neural dynamics,'' Nat. Phys. \textbf{6}, 744-750 (2010).
\bibitem{37} J. M. Beggs and D. Plenz, ``Neuronal avalanches in neocortical circuits,'' J. Neurosci. \textbf{23}, 11167-11177 (2003).


\end{thebibliography}
\end{document}